\documentclass[pre,preprint]{revtex4}
\begin{document}
\bibliographystyle{revtex4}
\titlepage
\title{A Model for the Coexistence of $p$-wave Superconductivity and
Ferroelectricity}
\author{Hong-Biao Zhang}
\email{zhanghb017@nenu.edu.cn} \affiliation{Institute of
Theoretical Physics, Northeast Normal University, Changchun
130024, P.R.China}
\author{Li-Jun Tian}
\affiliation{Department of Physics, College of Science, Shanghai
University, Shanghai 200436, P.R.China}
\author{Guo-Hong Yang}
\affiliation{Department of Physics, College of Science, Shanghai
University, Shanghai 200436, P.R.China}
\author{Shuo Jin}
\affiliation{Department of Physics, School of Science, Beihang
University, Beijing 100083, P.R.China}
\author{Mo-Lin Ge}
\affiliation{Theoretical Physics Division, Nankai Institute of
Mathematics, Nankai University, Tianjin 300071, P.R.China}
\begin{abstract}
 A model for the coexistence of $p$-wave superconductivity (SC)
 and ferroelectricity (FE) is presented. The Hamiltonian
 of SC sector and FE sector can be diagonalized
 by using the $so(5)$ and $h(4)$ algebraic coherent states respectively. We
 assume a minimal symmetry-allow coupling and simplify the
 total Hamiltonian through a double mean-field approximation (DMFA).
 A variational coherent-state (VCS) trial wave-function is applied
 for the ground state. It is found that the ferroelectricity gives
 rise to the magnetic field effect of $p$-wave superconductivity.
\end{abstract}
\maketitle \baselineskip=20pt
\section{Introduction}
The coexistence of superconductivity (SC) and ferroelectricity
(FE) and competition play a important role in motivating the work
of Bednorz and Muller \cite{muller} on the high temperature
cuprates superconductors. Even earlier, work on the ``old"
superconductors of the $\beta$-$W$ structure like $V_3Si$,
$Nb_3S_n$ etc., where $T_c\sim 23K$ and a martensitic phase
transition occurs in the same temperature range, gave rise to
investigations on the possibility of a ``ferroelectric metal" or a
``polar metal" and thus to the study of SC-FE coexistence
\cite{weger,anderson1,birman1,ting,bhatt}. Many theoretical papers
have already studied microscopic models for the effect of lattice
instability on superconductivity in the sodium tungsten bronze
systems \cite{ngai,ngai1,kahn}. These papers have illuminated many
aspects of the interplay between the structure deformations, such
as rotation of underlying octahedaral units and coupling with
electron pairs. Study of SC-FE coexistence problem can be relevant
to recent work by Weger and collaborators \cite{peter,peter1}, on
the mechanism of high temperature superconductivity in the
cuprates. In that work the presence of a nearby FE instability
close to the SC transition is related to the anomalously large
ionic dielectric coefficient in the cuprates, which reduces the
electron-electron repulsion and then can lead to an enhanced net
electron-electron attraction, producing higher $T_c$. Moreover the
recent $so(5)$ and $su(4)$ models for multi-critical
superconductor anti-ferromagnetic behavior in the high temperature
superconductors have also been studied by Zhang et.al.
\cite{zhang,zacher,zhang2,wu}. On the other hand, the recent
experimental discovery of the coexistence ferromagnestism (FM) and
SC in $VGe_{2}$ \cite{saxena,bauer}, and subsequently in
$ZrZn_{2}$ \cite{uhlarz} and $URhGe$ \cite{aoki}, have shown
clearly that the spin-triplet states of $p$-wave SC are realized
in the nature. This naturally causes our interest in the
relationship between FE and $p$-wave SC. This paper is organized
as follows: In Sec. II we will construct a $so(5)\otimes h(4)$
algebraic structure general model for the coexistence of $p$-wave
SC and FE. In Sec. III and IV we will diagonalize the Hamiltonian
of $p$-wave SC and FE according to $so(5)$ and $h(4)$ algebraic
coherent state methods respectively, and get their energies, the
eigenstates and the expection value of the order parameters in the
ground coherent state. In Sec. V the total Hamiltonian including
the biquadratic interaction is simplified to the bilinear
Hamiltonian $H_{DMFA}$ by using a double-mean-field approximation.
We can find the variational solution of $H_{DMFA}$ by forming a
trial eigenstate analogous to the product coherent state. The
energy spectrum and the eigenstates will be obtained based on the
variational solution.

\section{Our model}
Here we will consider a general model for the coexistence between
$p$-wave superconductivity and ferroelectricity. According to a
known result there is $so(5)$ structure in $p$-wave
superconductivity \cite{murakami,zhang1} that is formed by two
$su(2)$ not commuting with each other, where one describes the
attractive BCS interaction and the other the usual spin operators,
as well as other four generators associated with the transitions.
But ferroelectricity is formed by Heisenberg algebra $h(4)$
\cite{anderson,cochran,cowley,bruce}. Motivated by Ref.
\cite{birman}, we write the total Hamiltonian in three parts as
follows:
\begin{equation}
\label{H} H=H_{SC}+H_{FE}+H_{INT},
\end{equation}
where the first term $H_{SC}$ is the BW type of $p$-wave
superconductivity mean-field reduced Hamiltonian, namely
\begin{equation}
\label{HSC} H_{SC}=\sum_{\bf k} h_{\bf k},
\end{equation}
with the Hamiltonian at each given momentum $\bf k$ given as
\begin{eqnarray}
\label{h}
 h_{\bf k}&=&\epsilon_{\bf k}E_{3}^{(\bf k)}
+[\Delta_{E}({\bf k})E_+^{(\bf k)}+\Delta_{U}({\bf k})U_+^{(\bf
k)}+\Delta_{V}({\bf k})V_+^{(\bf k)} +H.c.].
\end{eqnarray}
Here $\Delta_{E}({\bf k})=\frac{1}{2}\sum\limits_{{\bf k}'}V_{{\bf
k}{\bf
 k}'}<E_{-}^{({\bf k}')}>$, $\Delta_{U}({\bf k})=\frac{1}{2}\sum\limits_{{\bf
k}'}V_{{\bf k}{\bf
 k}'}<U_{-}^{({\bf k}')}>$ and $\Delta_{V}({\bf
k})=\frac{1}{2}\sum\limits_{{\bf k}'}V_{{\bf k}{\bf
 k}'}<V_{-}^{({\bf k}')}>$ are opposite and equal spin pairing (``gap") energy
respectively,
 and $p$-wave attraction pair interaction potential $V_{{\bf k}{\bf
k}'}=-3 V_{1}(k,k')\bf k \cdot {\bf k}'$.

The second term $H_{FE}$ is the displaced oscillator Hamiltonian
of the displacive ferroelectric soft mode phonon \cite{landau}.
It's taken as
\begin{equation} \label{HFE} H_{FE}=\omega_{TO}(b^{\dag}b
+\frac{1}{2})+\gamma_{1}\varepsilon(b^{\dag}+b),
\end{equation}
where $\omega_{TO}$ is the frequency of the soft TO mode,
$\gamma_{1}$ is taken as a positive constant, and $\varepsilon$ is
the magnitude of the electric field $\vec{E}$.

The third term $H_{INT}$ is the interaction term. According to
Ginzburg-Landau (GL) theory \cite{landau1}, every term in the free
energy shall be a scalar invariant under the relevant symmetry
group, here the free energy density of system will be the form of
$\delta{F}=a|{\bf P}|^2+\frac{b}{4}|{\bf
P}|^4+\alpha|\Delta|^2+\beta|\Delta|^4+\kappa|\Delta|^2|{\bf
P}|^2$. In such a GL theory, the lowest-order, generic coupling
between superconducting $\Delta$ and the ferroelectric ${\bf P}$
will be a biquadratic term of the form proportional to
$|\Delta|^2{\bf P}^2$, that is, the last term in $\delta{F}$. This
lowest-order term satisfies gauge and parity symmetry
requirements. From the correspondences $|\Delta|^2\sim
(E_+E_-+U_+U_-+V_+V_-+H.c.)$ and ${\bf P}^2\sim (b+b^{\dag})^2$ we
can immediately translate this coupling term into the interaction
term of our model as follows:
\begin{equation} \label{HINT} H_{INT}=\sum_{\bf k}\gamma_{2\bf
k}(E_{+}^{(\bf k)}E_{-}^{(\bf k)}+U_{+}^{(\bf k)}U_{-}^{(\bf
k)}+V_{+}^{(\bf k)}V_{-}^{(\bf k)}+H.c.)(b^{\dag}+b)^2,
\end{equation}
where $\gamma_{2{\bf k}}$ is the initial pair-TO mode coupling
coefficient.

 In our model, the set $\{E_{\pm}^{(\bf
k)},E_{3}^{(\bf k)},F_{\pm}^{(\bf k)},F_{3}^{(\bf
k)},U_{\pm}^{(\bf k)},V_{\pm}^{(\bf k)}\}$ form $so(5)$ algebra,
its generators are expressed as
\begin{eqnarray}
\begin{array}{l}
\begin{array}{ll}
E_{+}^{(\bf k)}=\frac{1}{\sqrt{2}}(a^{\dag}_{-{\bf
k}\uparrow}a^{\dag}_{{\bf k}\downarrow}+a^{\dag}_{-{\bf
k}\downarrow}a^{\dag}_{{\bf k}\uparrow}), & F_{+}^{(\bf
k)}=\frac{1}{\sqrt{2}}(a^{\dag}_{{\bf k}\downarrow}a_{{\bf
k}\uparrow}+a^{\dag}_{-{\bf k}\downarrow}a_{-{\bf k}\uparrow}),\cr
E_{-}^{(\bf k)}=\frac{1}{\sqrt{2}}(a_{{\bf k}\downarrow}a_{-{\bf
k}\uparrow}+a_{{\bf k}\uparrow}a_{-{\bf k}\downarrow}), &
F_{-}^{(\bf k)}=\frac{1}{\sqrt{2}}(a^{\dag}_{{\bf
k}\uparrow}a_{{\bf k}\downarrow}+a^{\dag}_{-{\bf
k}\uparrow}a_{-{\bf k}\downarrow}), \cr U_{+}^{(\bf
k)}=a^{\dag}_{-{\bf k}\uparrow}a^{\dag}_{{\bf
k}\uparrow},&U_{-}^{(\bf k)}=a_{{\bf k}\uparrow}a_{-{\bf
k}\uparrow}, \cr V_{+}^{(\bf k)}=a^{\dag}_{-{\bf
k}\downarrow}a^{\dag}_{{\bf k}\downarrow},&V_{-}^{(\bf k)}=a_{{\bf
k}\downarrow}a_{-{\bf k}\downarrow},
\end{array}
\cr E^{{\bf k}}_3=\frac{1}{2}(a^{\dag}_{{\bf k}\uparrow}a_{{\bf
k}\uparrow}+a^{\dag}_{-{\bf k}\uparrow}a_{-{\bf
k}\uparrow}+a^{\dag}_{{\bf k}\downarrow}a_{{\bf
k}\downarrow}+a^{\dag}_{-{\bf k}\downarrow}a_{-{\bf
k}\downarrow}-2), \cr F^{{\bf k}}_3=\frac{1}{2}(a^{\dag}_{{\bf
k}\downarrow}a_{{\bf k}\downarrow}+a^{\dag}_{-{\bf
k}\downarrow}a_{-{\bf k}\downarrow}-a^{\dag}_{{\bf
k}\uparrow}a_{{\bf k}\uparrow}-a^{\dag}_{-{\bf k}\uparrow}a_{-{\bf
k}\uparrow}),
\end{array}
\end{eqnarray}
and satisfy the following commutation relations:
\begin{eqnarray}
\label{e6}
\begin{array}{l}
\begin{array}{lll}
[E_{\pm}^{(\bf k)}, V_{\mp}^{(\bf k)}]=\pm F_{\mp}^{(\bf k)},&
\;\;[F_{\pm}^{(\bf k)}, V_{\mp}^{(\bf k)}]=\mp E_{\mp}^{(\bf k)},&
\;\;[E_{\pm}^{(\bf k)},U_{\mp}^{(\bf k)}]=\pm F_{\pm}^{(\bf
k)},\cr [F_{\pm}^{(\bf k)}, U_{\pm}^{(\bf k)}]=\pm E_{\pm}^{(\bf
k)},& \;\;[E_{\pm}^{(\bf k)}, F_{\pm}^{(\bf k)}]=\mp V_{\pm}^{(\bf
k)},& \;\;[E_{\pm}^{(\bf k)}, F_{\mp}^{(\bf k)}]=\mp U_{\pm}^{(\bf
k)},\cr [E_{3}^{(\bf k)}, E_{\pm}^{(\bf k)}]=\pm E_{\pm}^{(\bf
k)},& \;\;[F_{3}^{(\bf k)}, F_{\pm}^{(\bf k)}]=\pm F_{\pm}^{(\bf
k)},& \;\;[E_{3}^{(\bf k)}, U_{\pm}^{(\bf k)}]=\pm U_{\pm}^{(\bf
k)},\cr [F_{3}^{(\bf k)}, U_{\pm}^{(\bf k)}]=\mp U_{\pm}^{(\bf
k)},& \;\;[E_{3}^{(\bf k)}, V_{\pm}^{(\bf k)}]=\pm V_{\pm}^{(\bf
k)},& \;\;[F_{3}^{(\bf k)}, V_{\pm}^{(\bf k)}]=\pm V_{\pm}^{(\bf
k)},
\end{array}\cr
\begin{array}{ll}
[E_{+}^{(\bf k)}, E_{-}^{(\bf k)}]=E_{3}^{(\bf k)},\;\;\;\;&
[U_{+}^{(\bf k)}, U_{-}^{(\bf k)}]=(E_{3}^{(\bf k)}-F_{3}^{(\bf
k)}), \cr [F_{+}^{(\bf k)}, F_{-}^{(\bf k)}]=F_{3}^{(\bf
k)},\;\;\;\;& [V_{+}^{(\bf k)}, V_{-}^{(\bf k)}]=(E_{3}^{(\bf
k)}+F_{3}^{(\bf k)}).
\end{array}
\end{array}
\end{eqnarray}
Note that $H_{SC}$ is linearized with respect to $so(5)$ algebraic
generators for a given momentum ${\bf k}$, $H_{FE}$ is written in
terms of $h(4)$ generators, and $H_{INT}$ is expressed as a
product of the bilinear form of so(5) and quadratic form of
$h(4)$. Therefore the total Hamiltonian is a $so(5)\otimes h(4)$
direct product algebraic structure in each given momentum ${\bf
k}$.
\section{The $so(5)$ structure of $p$-wave superconductivity model}
From Hamiltonian $H_{SC}$ we can known that $h_{\bf k}$ is written
in terms of $so(5)$ generators for a given $\bf k$. The dynamical
symmetry or spectrum generating algebra for each ${\bf k}$ is
$so(5)_{\bf k}$, so the spectrum generating algebra of $H_{SC}$ is
${\otimes}_{\bf k}so(5)_{\bf k}$.

The eigenstates of $H_{SC}$ are expressed by a direct product of
$so(5)_{\bf k}$ coherent states $\otimes_{\bf k} |\xi_{\bf k}
\rangle$. Therefore the eigenstates $|\xi \rangle$ can be written
as
\begin{eqnarray}
\label{e12} |\xi \rangle=\otimes_{\bf k}|\xi_{\bf k} \rangle
            =\otimes_{\bf k}W(\xi_{\bf k})|p,q>,
\end{eqnarray}
where
\begin{eqnarray*}
 W(\xi_{\bf k})&=&\exp\{\xi_{\bf k}(\sqrt{2}\cos{\theta_{\bf k}}E^{(\bf
k)}_{+}
  -\sin{\theta_{\bf k}}e^{i\phi_{\bf k}}U^{(\bf k)}_{+} \\
& & +\sin{\theta_{\bf k}}e^{-i\phi_{\bf k}}V^{(\bf k)}_{+})
      -H.C.\},
\end{eqnarray*}
with the real coherent parameters $ \xi_{\bf k}$, $(\theta_{\bf
k},\phi_{\bf k})$ are angles in spin space for a given momentum
$\bf k$. $|p,q \rangle$ is the mutual eigenstates of the Cartan
subalgebra $\{E^{(\bf k)}_3,F^{(\bf k)}_3\}$ of $so(5)$, and their
eigenvalues are $p$ and $q$ respectively. By tedious calculations
we can immediately diagonalize the Hamiltonian $h_{\bf k}$ as
\begin{eqnarray} \label{e13}
 W^{\dag}(\xi_{\bf k})h_{\bf k}W(\xi_{\bf k})
 &=&\sqrt{\epsilon^2_{\bf k}+\Delta^2({\bf k})}E^{(\bf k)}_{3},
 \end{eqnarray}
where $\Delta^2({\bf k})=2(|\Delta_{E}({\bf
k})|^2+|\Delta_{U}({\bf k})|^2+|\Delta_{V}({\bf k})|^2)$. Also we
obtain the gap equation:
\begin{eqnarray}
\label{gap} \left[\begin{array}{l} \Delta_{E}({\bf k})\cr
\Delta_{U}({\bf k})\cr\Delta_{V}({\bf
k})\end{array}\right]=\frac{1}{2}\sum_{{\bf k}'}V_{{\bf k}{\bf
k}'}\frac{\Delta({\bf k}')}{\sqrt{\epsilon^2_{{\bf
k}'}+\Delta^2({\bf k}')}}
\left[\begin{array}{l}\frac{-p}{\sqrt{2}}\cos{\theta_{{\bf
k}'}}\cr \frac{p-q}{2}\sin{\theta_{{\bf k}'}}e^{i\phi_{{\bf
k}'}}\cr\frac{-(p+q)}{2}\sin{\theta_{{\bf k}'}}e^{-i\phi_{{\bf
k}'}}\end{array}\right].
\end{eqnarray}

From the above gap equation we can obtain the following results:

1. If both $p$ and $q$ are zero, the gap $\Delta=0$ and the
$p$-wave SC lies in disorder state. Its eigenstate is
$|\xi_{dis}>=\otimes_{\bf k}W(\xi_{\bf k})|p=0,q=0>$.

2. If either $p$ or $q$ are not zero, the gap $\Delta\not=0$ and
the $p$-wave SC lies in superconducting state, but when $p=-1$ and
$q=0$ the $p$-wave SC lies in the ground state. Its ground state
is $|\xi_{BCS}>=\otimes_{\bf k}W(\xi_{\bf k})|p=-1,q=0>$ where
$|p=-1,q=0>$ is the vacuum state.
\section{A $h(4)$ structure of displacive ferroelectric soft mode}
The Hamiltonian $H_{FE}$ can be transformed by the displaces
oscillator Bose operator
\begin{equation}
\label{u} U(\zeta_{0})=\exp[\zeta_{0}(b^{\dag}-b)],
\end{equation}
where the coherent parameter $\zeta_{0}$ is taken as real, so we
obtain
\begin{eqnarray}
U^{\dag}H_{FE}U=\omega_{TO}(b^{\dag}b+\frac{1}{2})
-\frac{(\gamma_{1}\varepsilon)^2}{\omega_{TO}}.
\end{eqnarray}
Note that $U^{\dag}H_{FE}U$ is shifted to a new minimum, but it
retains the same excitation frequency $\omega_{TO}$ as the
original oscillator. The eigenstates and eigenvalues of $H_{FE}$
can be given as
\begin{equation}
|\zeta_{0}>=U(\zeta_{0})|n>=\exp[\zeta_{0}(b^{\dag}-b)]|n>,
\end{equation}
and
\begin{equation}
W_{n}=\omega_{TO}(n+\frac{1}{2})-\frac{(\gamma_{1}\varepsilon)^2}{\omega_{TO}}
,
\end{equation}
where $|n>$ is a number eigenstate of the phonon number operator
$N_{b}=b^{\dag}b$ and the state $|\zeta_{0}>$ is a Glauber
coherent state for the FE oscillator \cite{klauder,glauber}.

The order operator for the FE polarization is the coordinate
operator $Q$ or $(b+b^{\dag})$, therefore the order parameter
$$\eta_{FE}=\left\{\begin{array}{l}<\zeta_{0}|b^{\dag}+b|\zeta_{0}>=2\zeta_{0}

\cr <n|b^{\dag}+b|n>=0.\end{array}\right. $$ When $\eta_{FE}=0$ it
shows that the phonon is free, but $\eta_{FE}=2\zeta_{0}$ it
exhibits that the phonon is polarized spontaneously.

\section{The variational coherent state eigenstates of our
Hamiltonian under double-mean-approximation} The total Hamiltonian
is turned into bilinear forms by double mean field approximation
procedure which reduces biquadratic operators such as $A^2B^2$ to
the following form
\begin{eqnarray}
A^2B^2&\approx&(2A<A>-<A>^2)(2B<B>-<B>^2),
\end{eqnarray}
based on the assumption $(A-<A>)^2\approx 0$ and $(B-<B>)^2\approx
0$. After making the double-mean-approximation and isolating a
single mode ${\bf k}$, we have the effective Hamiltonian at each
momentum mode ${\bf k}$ in the DMFA:
\begin{eqnarray}
\label{Heff}
 H_{DMFA}&=&\epsilon{E}_{3}
+(\Delta^{'}_{E}E_++\Delta^{'}_{U}U_++\Delta^{'}_{V}V_++H.c.)
+\omega_{TO}(b^{\dag}b+\frac{1}{2})+\Gamma_{1}(b^{\dag}+b)
\nonumber \\
&&+(\Gamma^{E}_2E_++\Gamma^{U}_2U_++\Gamma^{V}_2V_++H.c.)(b^{\dag}+b)+\Gamma_{3}.
\end{eqnarray}
Therefore the double mean field approximation (DMFA) yields a
bilinear effective interaction term, and it renormalizes the
coefficients $\Delta^{'}$ and $\Gamma_{1,2,3}$ in $H_{DMFA}$ as
follows:
\begin{equation}
\label{coefficient}
\begin{array}{l}
\begin{array}{ll}
\left\{
\begin{array}{l}\Delta^{'}_{E}
=\Delta_{E}-\gamma_{2}<b^{\dag}+b>^2<E_->, \cr
\Delta^{'}_{U}=\Delta_{U}-\gamma_{2}<b^{\dag}+b>^2<U_->, \cr
\Delta^{'}_{V}=\Delta_{V}-\gamma_{2}<b^{\dag}+b>^2<V_->,
\end{array}\right.
& \left\{\begin{array}{l}
\Gamma^{E}_2=2\gamma_{2}<b^{\dag}+b><E_->,\cr
\Gamma^{U}_2=2\gamma_2<b^{\dag}+b><U_->, \cr
\Gamma^{V}_2=2\gamma_2<b^{\dag}+b><V_->,
\end{array}\right.
\end{array}
\cr
\Gamma_{1}=\gamma_{1}\varepsilon-2\gamma_{2}<b^{\dag}+b>(<E_+><E_->
+ <U_+><U_->+<V_+><V_->), \cr \Gamma_{3}=\gamma_{2}(<E_+><E_-> +
<U_+><U_-> + <V_+><V_->)<b^{\dag}+b>^2.
\end{array}
\end{equation}
This total Hamiltonian includes the pure $p$-wave SC and the pure
FE prototype systems and their coupling via the soft-mode
oscillator coupled to opposite and equal-spin pairing Hamiltonian.
Noting that the initial Hamiltonian is the enveloping algebra of
$so(5)\otimes h(4)$ because of the biquadratic interaction terms,
while $H_{DM FM}$ is an element in the direct product algebra
$so(5)\otimes h(4)$.

When $\gamma_{2}\rightarrow 0$, we recover the sum of two separate
sectors: for $p$-wave SC and FE. In order to obtain the ground
state eigenstates and eigenvalue of our model $H_{DMFA}$, we will
make use of the variational principle. Introducing an analogous
trial variational coherent state (VCS) which is the product of two
coherent-like states and is denoted $|\varphi_{\nu}>$:
\begin{eqnarray}
\label{}
|\varphi_{\nu}>=|\xi>|\zeta>=\hat{V}_{1}|p,q>\hat{V}_{2}|n>,
\end{eqnarray}
where
$V_{1}(\xi)=\exp[\xi(\sqrt{2}\cos{\theta}E_+-\sin{\theta}e^{i\phi}U_+
 +\sin{\theta}e^{-i\phi}V_+)-H.c.]$ and
$V_{2}(\zeta)=\exp[\zeta(b^{\dag}-b)]$. Here the real parameters
$\xi$ and $\zeta$ are variational unknowns. The kets $|p,q>$ and
$|n>$ are the same as before. Now we define the energy in state
$|\varphi_{\nu}>$ as the diagonal value of $H_{DMFA}$ in the
variational coherent state $|\varphi_{\nu}>$:
\begin{eqnarray}
\label{xy}
&&E_{p,q,n}(\xi,\theta,\phi,\zeta)=<\varphi_{\nu}|H_{DMFA}|\varphi_{\nu}>
\nonumber \\
&=&p\epsilon\cos{2\xi}+[-\frac{p}{\sqrt{2}}\Delta^{'}_{E}\cos{\theta}
+\frac{p-q}{2}\Delta^{'}_{U}e^{-i\phi}\sin{\theta}
-\frac{p+q}{2}\Delta^{'}_{V}e^{i\phi}\sin{\theta}+H.c.]\sin{(2\xi)}
\nonumber \\
&&+2\zeta[-\frac{p}{\sqrt{2}}\Gamma^{E}_2\cos{\theta}
+\frac{p-q}{2}\Gamma^{U}_2e^{-i\phi}\sin{\theta}
-\frac{p+q}{2}\Gamma^{V}_2e^{i\phi}\sin{\theta}+H.c.]\sin{(2\xi)}
 \nonumber \\
&&+\omega_{TO}(n+\zeta^2+\frac{1}{2})+2\Gamma_1\zeta +\Gamma_3
\nonumber \\
&=&p\epsilon\cos{(2\xi)}+p\Delta\sin{(2\xi)}
+\omega_{TO}(n+\zeta^2+\frac{1}{2})
\nonumber \\
&&+6\gamma_2\zeta^2(p^2+q^2\sin^{2}{\theta})\sin^{2}{(2\xi)}
+2\gamma_1\varepsilon\zeta.
\end{eqnarray}
Here the coefficients $\Delta^{'}$ and $\Gamma_{1,2,3}$ in the
energy $E_{p,q,n}(\xi,\theta,\phi,\zeta)$ are given as
\begin{equation}
\begin{array}{l}
\left\{\begin{array}{l}
\Delta^{'}_{E}=\Delta_E-4\gamma_2\zeta^2\sin{(2\xi)}(-\frac{p}{\sqrt{2}}\cos{\
theta}) =-\frac{1}{\sqrt{2}}\cos{\theta}
[\Delta-4\gamma_2\zeta^2p\sin{(2\xi)}],\cr
\Delta^{'}_{U}=\Delta_U-4\gamma_2\zeta^2\sin{(2\xi)}(\frac{p-q}{2}\sin{\theta}
e^{i\phi})
=\frac{1}{2}\sin{\theta}e^{i\phi}[\Delta-4\gamma_2\zeta^2(p-q)\sin{(2\xi)}]
,\cr
\Delta^{'}_{V}=\Delta_V-4\gamma_2\zeta^2\sin{(2\xi)}(-\frac{p+q}{2}\sin{\theta
}e^{-i\phi})
=-\frac{1}{2}\sin{\theta}e^{-i\phi}[\Delta-4\gamma_2\zeta^2(p+q)\sin{(2\xi)}],
\end{array}\right.
\cr \left[\begin{array}{l}\Gamma^{E}_2\cr \Gamma^{U}_2 \cr
\Gamma^{V}_2
\end{array}\right]
=4\gamma_2\zeta\sin{(2\xi)}
\left[\begin{array}{l}-\frac{p}{\sqrt{2}}\cos{\theta}\cr
\frac{p-q}{2}\sin{\theta}e^{i\phi}\cr-\frac{p+q}{2}\sin{\theta}e^{-i\phi}
\end{array}\right]
, \cr
\Gamma_1=\gamma_1\varepsilon-2\gamma_2\zeta(p^2+q^2\sin^2{\theta})\sin^2{(2\xi
)}, \cr
\Gamma_3=2\gamma_2\zeta^2(p^2+q^2\sin^2{\theta})\sin^2{(2\xi)}.
\end{array}
\end{equation}

We determine $\theta,\phi,\xi,\zeta$ from the following equations:
\begin{equation}
\label{gh} \left\{
\begin{array}{l}
\partial{E}/\partial{\theta}=0,\cr
\partial{E}/\partial{\phi}=0,\cr
\partial{E}/\partial{\xi}=0,\cr
\partial{E}/\partial{\zeta}=0.
\end{array}
\right.
\end{equation}
Eqs. (\ref{gh}) have two solutions:

(i).  $\left\{
\begin{array}{l}
\sin{\theta}=0, \cr
\tan{2\xi}=[\Delta+12\gamma_2\zeta^2{p}\sin{(2\xi)}]/{\epsilon},
\cr
\xi=-[\gamma_1\varepsilon+6\gamma_2\zeta{p^2}\sin^2{(2\xi)}]/{\omega_{TO}}.
\end{array}\right.$

This result exhibits that FE and SC can coexist in the opposite
spin pairing state of $p$-wave SC but the equal spin pairing state
of $p$-wave SC disappeared. Here the energy is
$E=p\epsilon\cos{(2\xi)}+p\Delta\sin{(2\xi)}
+\omega_{TO}(n+\zeta^2+\frac{1}{2})+6\gamma_2\zeta^2p^2\sin^2{(2\xi)}
+2\gamma_1\varepsilon\zeta$.

(ii).  $\left\{\begin{array}{l} \cos{\theta}=0, \cr
\tan{2\xi}=[\Delta+12\gamma_2\zeta^2(p+\frac{q^2}{p})\sin{(2\xi)}]/{\epsilon},
\cr\xi=-[\gamma_1\varepsilon+6\gamma_2\zeta(p^2+q^2)\sin^2{(2\xi)}]/{\omega_{T
O}}.
\end{array}\right.$

This result exhibits that FE and $p$-wave SC may coexist in the
equal spin pairing state (ABM state) but the opposite spin pairing
state of $p$-wave SC disappeared. Here the energy is
$E=p\epsilon\cos{(2\xi)}+p\Delta\sin{(2\xi)}
+\omega_{TO}(n+\zeta^2+\frac{1}{2})+6\gamma_2\zeta^2(p^2+q^2)\sin^2{(2\xi)}
+2\gamma_1\varepsilon\zeta$.

Therefore from above two cases we get that the spin of $p$-wave SC
is oriented because of the existence of FE. It indicates that the
FE effect in the excite state of $p$-wave SC is equivalent to
magnetic field.

\section{Conclusion}
In this paper we have constructed a general model for the
coexistence of $p$-wave superconductivity and ferroelectricity.
The Hamiltonian of $p$-wave SC can been diagonalized based on
$so(5)$ spectrum-generating algebra structure and has shown that
the eigenstate is related to $so(5)$ coherent state. Also the
Hamiltonian of FE is diagonalized by using $h(4)$ algebraic
coherent state. The total Hamiltonian under the double mean-field
approximation can been solved by making use of a variational
coherent-state (VCS) procedure by virtue of the $so(5)\otimes
h(4)$ algebraic structure in each given momentum ${\bf k}$. This
leads to the conclusion that the ferroelectricity gives rise to
the magnetic field effect of $p$-wave superconductivity.

\vspace{3mm}

\section*{Acknowledgements}

This work is supported in part by the NSF of China under Grant
No.10405006 and 10201015. \vspace{3mm}

\noindent
\end{document}